\documentclass[preprint,prd,aps,nofootinbib,onecolumn,11pt]{revtex4}

\usepackage{amsmath}
\usepackage{epsfig}
\usepackage{amssymb}
\usepackage{color}

\begin{document}

\def\rd{\rm d}

\newcommand{\RM}{\ensuremath{\mathrm}}   
\newcommand{\DD}{\mathrm{d}}  
\newcommand{\me}{\mathrm{e}}  
\newcommand{\mi}{\mathrm{i}}  
\newcommand{\mj}{\mathrm{j}}  
\newcommand{\afrac}[2]{\dfrac{\,#1\,}{\,#2\,}}  
\newcommand{\nn}{\nonumber}  
\newcommand{\nt}{\noindent}
\newcommand{\LT}{\left}
\newcommand{\RT}{\right}

\newcommand{\blue}[1]{\textcolor{blue}{#1}}
\newcommand{\red}[1]{\textcolor{red}{#1}}
\newcommand{\kk}{$K_1(1270)$ and $K_1(1400)$}
\newcommand{\ka}{$K_1(1270)$}
\newcommand{\kb}{$K_1(1400)$}


\title{Strange Axial-vector Mesons in $D$ Meson Decays} %

\author{Peng-Fei Guo, Di Wang, Fu-Sheng Yu}

\affiliation{School of Nuclear Science and Technology,  Lanzhou University,
Lanzhou 730000, China}

\begin{abstract}
The nature of strange axial-vector mesons are not well understood and can be investigated in $D$ meson decays.
In this work, it is found that the experimental data of $D^0\rightarrow K^\pm K^\mp_1(1270)(\to \rho K \,\,\text{or}\,\,K^{*}\pi)$ in the $D^0\to K^+K^-\pi^+\pi^-$  mode, disagree with the equality relation under the narrow width approximation and $CP$ conservation of strong decays.
Considering more other results of $K_1(1270)$ decays, the data of $\mathcal{B}(D^0\rightarrow K^-K^+_1(1270)(\to K^{*0}\pi^+))$ is probably overestimated by one order of magnitude.
We then calculate the branching fractions of the corresponding processes with $K_1(1400)$ in the factorization approach, and find $\mathcal{B}(D^0\rightarrow K^-K^+_1(1400)(\to K^{*0}\pi^+))$ is comparable to the predicted $\mathcal{B}(D^0\rightarrow K^-K^+_1(1270)(\to K^{*0}\pi^+))$ using the equality relation.
Besides, we suggest to measure the ratios between $K_1(1270)\to \rho K$ and $K^*\pi$  or to test the equality relations in other $D$ meson decay modes.
\end{abstract}

\maketitle

\section{Introduction}   

In the quark model, there are two nonets of axial-vector ($J^P=1^+$) mesons, namely, $^3P_1$ and $^1P_1$ in the spectroscopic notation $^{2S+1}L_J$, which correspond to the charge parity of $C=+$ and $C=-$, respectively, for the neutral mesons with isospin $I_3=0$ in each nonet.
The strange axial-vector mesons in these two nonets are called as $K_{1A}$ and $K_{1B}$, respectively.
They can mix with each other to construct the mass eigenstates, $K_1(1270)$ and $K_1(1400)$, by the mixing angle $\theta_{K_1}$:
\begin{align}\label{mixing}
\begin{pmatrix}
|K_1(1270)\rangle   \\
|K_1(1400)\rangle   \\
\end{pmatrix}
=
\begin{pmatrix}
\sin\theta_{K_1}        &       \cos\theta_{K_1}        \\
\cos\theta_{K_1}        &       -\sin\theta_{K_1}       \\
\end{pmatrix}
\begin{pmatrix}
|K_{1A}\rangle      \\
|K_{1B}\rangle      \\
\end{pmatrix}.
\end{align}
The experimental measurements on $K_1(1270)$ and $K_1(1400)$ have been performed in $Kp$ scattering \cite{Daum:1981hb,Aston:1986jb}, $\tau^\pm$ decays \cite{Bauer:1993wn,Barate:1999hj,Abbiendi:1999cq,Asner:2000nx}, $B$-meson decays \cite{Abe:2001wa,Yang:2004as,Aubert:2008bc,Aubert:2009ab,Guler:2010if,Sanchez:2015pxu} and $D$-meson decays \cite{Aitala:1997gy,Link:2004wx,Artuso:2012df,1703.08505,Ablikim:2017eqz,Aaij:2017kbo}. However, the mixing angle  $\theta_{K_1}$ has not yet been well determined.
Many phenomenological analysis indicate that the value of $\theta_{K_1}$ is around either $35^\circ$ or $55^\circ$ through the strong decays of $K_1(1270)$ and $K_1(1400)$
\cite{Suzuki:1993yc,Divotgey:2013jba},
$\tau\rightarrow K_1(1270),K_1(1400)\nu$ \cite{Suzuki:1993yc}, $B\rightarrow K_1(1270),K_1(1400)\gamma$ \cite{Hatanaka:2008xj} and the mass relation \cite{Cheng:2011pb}, $\theta_{K_1}\sim45^\circ$ in the relativized quark model \cite{Blundell:1995au} and the modified Godfrey-Isgur model \cite{1705.03144}, or $\theta_{K_1}\sim60^\circ$ based on the $^3P_0$ quark-pair-creation model for decays of $K_1(1270)$ and $K_1(1400)$ \cite{Tayduganov:2011ui}.
 $35^\circ\lesssim\theta_{K_1}\lesssim65^\circ$ are obtained in some other analysis \cite{Godfrey:1985xj,Lipkin:1977uy,Burakovsky:1997dd}.

The mixing angle $\theta_{K_1}$ can also be investigated in heavy flavor decays. The difference between the production rates of \kk ~may provide the indication on the value of $\theta_{K_1}$. It has been widely studied in $B$-meson decays, such as hadronic decays of $B\to K_1(1270),K_1(1400) P(V)$ \cite{Cheng:2009ms,Cheng:2003sm,Chen:2005cx,Calderon:2007nw,Cheng:2008gxa,Liu:2010da,Liu:2010nz,Liu:2014dxa,Zhang:2017cbi,Sayahi:2012zz}, with $P=\pi, K, \eta^{(\prime)}$, and $V=\rho, \omega, K^*, \phi, J/\Psi $, semi-leptonic decays of $B\to K_1(1270),K_1(1400)\ell^+\ell^-$ \cite{Choudhury:2007kx,Bayar:2008ug,Hatanaka:2008gu,Bashiry:2009wq,Li:2009rc}, and radiative decays of $B\to K_1(1270),K_1(1400)\gamma$ \cite{Cheng:2004yj,Lee:2006qj,Hatanaka:2008xj,Kou:2010kn}.
The two-body hadronic $D$-meson decays with an axial-vector meson in the final states have been studied in \cite{Cheng:2003bn,Kamal:1991kg,Pham:1991az,Pham:1991ex,Kamal:1993gu,Katoch:1995bm,Lipkin:2000gz,Cheng:2010vk}.
The large non-perturbative contributions in charm decays always pollute the analysis on the $K_1(1270)$ and $K_1(1400)$ productions.
On the other hand, at the LHCb, more data of $D$ decays are obtained than $B$ decays, due to the larger production cross sections of $D$ mesons and the larger branching fractions of $D$ decays.
Besides, the running BESIII and the upcoming Belle II experiments will provide large data of $D$ decays as well.
For example, $K_1(1270)$ and $K_1(1400)$ have been analyzed in the $D^0\to K^-\pi^+\pi^+\pi^-$ mode at the BESIII \cite{Ablikim:2017eqz} and LHCb \cite{Aaij:2017kbo} very recently.
With the large data and thus high precision of measurements in the near future, the processes of $D$ decaying into $K_1(1270)$ and $K_1(1400)$ are worthwhile to be studied with more efforts.

Among the exclusive $D\to K_1(1270),K_1(1400)$ decays, the $D^0\to K^+K^-\pi^+\pi^-$ mode is of particular interest since there are more cascade channels involving $K^-K_1^+(1270)(\to K^+\rho^0(\to \pi^+\pi^-))$, $K^-K_1(1270)^+(\to \pi^+K^{*0}(\to K^+\pi^-))$, $K^+K_1(1270)^-(\to K^-\rho^0(\to \pi^+\pi^-))$, $K^+K_1(1270)^-(\to \pi^-\overline K^{*0}(\to \overline{K}^0\pi^+))$, and the corresponding ones with $K_1^\pm(1400)$ instead of $K_1^\pm(1270)$. Besides, all the particles in the final states are charged and thus easier to be measured in experiments.
So far the relevant measurements have been performed by the E791 \cite{Aitala:1997gy}, FOCUS \cite{Link:2004wx} and CLEO \cite{Artuso:2012df} collaborations.
In \cite{Artuso:2012df}, only $K_1^\pm(1270)$ are involved but with $K_1^\pm(1400)$ neglected.
The fractions of decay widths of $D^0\to K^\pm K_1^\mp(1270)(\to \rho K, K^*\pi\to K^\mp\pi^\pm\pi^\mp)$ compared to that of $D^0\to K^+K^-\pi^+\pi^-$ are shown in Table\ 1.
\footnote{Very recently, PDG\cite{Patrignani:2016xqp} reversed these decay modes according to the re-analysis on the CLEO data by \cite{1703.08505}. We will discuss on it in Sec.4.}

\begin{table}
\caption{
List of the fractions for the $K_1^\pm(1270)$-involved cascade modes in the $D^0\rightarrow K^+K^-\pi^+\pi^-$ decay measured by CLEO \cite{Artuso:2012df}, $\Gamma(D^0\to K^\pm K_1^\mp(1270)(\to \rho K, K^*\pi\to K^\mp\pi^\pm\pi^\mp))/\Gamma(D^0\rightarrow K^+K^-\pi^+\pi^-)$. The first and second
uncertainties are statistical and systematic respectively.}
\begin{tabular}{ccccc}
\hline\hline

Modes& ~~Fractions ($\%$)\\

\hline

$K^-K_1(1270)^+(\to \pi^+K^{*0}(\to K^+\pi^-))$&   $7.3\pm0.8\pm1.9$    \\
\hline
$K^+K_1(1270)^-(\to K^-\rho^0(\to \pi^+\pi^-))$&   $6.0\pm0.8\pm0.6$   \\
\hline
$K^-K_1^+(1270)(\to K^+\rho^0(\to \pi^+\pi^-))$&   $4.7\pm0.7\pm0.8$   \\
\hline
$K^+K_1(1270)^-(\to \pi^-\overline K^{*0}(\to K^-\pi^+))$&   $0.9\pm0.3\pm0.4$  \\
\hline
\end{tabular}
\end{table}

We find a puzzle in the fractions given in Table\ 1 .
In the narrow width approximation and the $CP$ conservation of strong decays, the four partial widths satisfy a relation of
\begin{align}\label{eq:puzzle}
&\frac{\Gamma(D^0\to K^-K^+_1(1270), ~K^+_1(1270)\rightarrow K^{*0}\pi^+)}{\Gamma(D^0\to K^-K^+_1(1270),~K^+_1(1270)\rightarrow \rho^0 K^+)}\nonumber\\
=&\frac{\Gamma(D^0\to K^+K^-_1(1270),~K^-_1(1270)\rightarrow \overline{K}^{*0}\pi^-)}{\Gamma(D^0\to K^+K^-_1(1270),~K^-_1(1270)\rightarrow \rho^0 K^-)},
\end{align}
in which the weak-decay parts are canceled and it retains only the strong decays of the $K_1(1270)$.
However, from Table\ 1, the left-hand side of the above relation is 1.55$\pm$0.56, while the right-hand side is 0.15$\pm$0.09. They deviate from the equality relation by more than $2\sigma$. The central values are even different by a factor of 10.

We calculate the branching fractions of $D^0\to K^\pm K_1^\mp(1400)$ considering the finite-width effect in the factorization approach. It is found that the branching fraction of $D^0\rightarrow K^-K^+_1(1400), K_1^+(1400)\to K^{*0}\pi^+, K^{*0}\to K^+\pi^-)$ is comparable to $D^0\rightarrow K^-K^+_1(1270),K^+_1(1270)\to K^{*0}\pi^+, K^{*0}\to K^+\pi^-)$. Thus the inclusion of  $K_1(1400)$ in $1^+$ state may contribute to the overestimation of the latter process. Besides, we propose to test some relations of $D$ mesons decaying into $K_1(1270)$ processes in the subsequent measurements.

This paper is organized as follows.
In Sec.~2, we discuss the puzzle of the experimental data of $D^0\rightarrow K^+K^-\pi^+\pi^-$ decays with $K_1(1270)$ resonances.
In Sec.~3, the branching fractions of $D\to K_1(1400)$ transitions are estimated.
Some relations about $D$ decays into $K_1(1270)$ are listed in Sec.~4. And Sec.~5 is the conclusion.

\section{$K_1$ puzzle in $D^0\rightarrow K^+K^-\pi^+\pi^-$}\label{exp}

The puzzle introduced above is based on the narrow width approximation in the chain decays of heavy mesons. Taking the process of $D\to f_1f_2f_3$ with a resonant contribution of $R\to f_2f_3$ as an example, the branching fraction of $D\rightarrow f_1 R\to f_1f_2f_3$ is the product of branching fractions of $D\rightarrow f_1R$ and $R\rightarrow f_2f_3$:
\begin{align}\label{app}
\mathcal{B}(D\rightarrow f_1 R\to f_1f_2f_3)=\mathcal{B}(D\rightarrow f_1R)\,\mathcal{B}(R\rightarrow f_2f_3).
\end{align}
The narrow width approximation is valid in the decay of $D\to KK_1(1270), K_1(1270)\to K\pi\pi$ where the first decay is kinematically allowed and the width of $K_1(1270)$ is much smaller than its mass, $\Gamma_{K_1(1270)}\ll m_{K_1(1270)}$, as seen in Table\ 2.

\begin{table}
\caption{
Masses and widths of \kk. Data are from PDG \cite{Patrignani:2016xqp}.
}\begin{tabular}{cccc}

\hline\hline

~ & Mass & Width \\

\hline

$K_1(1270)$ & ~~$1272\pm7$ MeV~~ & $90\pm20$ MeV\\
\hline
$K_1(1400)$ & ~~$1403\pm7$ MeV~~ & $174\pm13$ MeV\\

\hline\hline
\label{tab:K1}
\end{tabular}\\
\end{table}

Therefore, the ratios of branching fractions of the processes in Eq. (\ref{eq:puzzle}) are thus
\begin{align}\label{eq:p1}
&{\mathcal{B}(D^0\to K^-K^+_1(1270), K^+_1(1270)\rightarrow K^{*0}\pi^+)
\over
\mathcal{B}(D^0\to K^-K^+_1(1270), K^+_1(1270)\rightarrow \rho^0 K^+)}\nonumber\\
=&
{\mathcal{B}(D^0\to K^-K^+_1(1270))~\mathcal{B}(K^+_1(1270)\rightarrow K^{*0}\pi^+)
\over
\mathcal{B}(D^0\to K^-K^+_1(1270))~\mathcal{B}(K^+_1(1270)\rightarrow \rho^0 K^+)}
\nonumber\\=&
{\mathcal{B}(K^+_1(1270)\rightarrow K^{*0}\pi^+)
\over
\mathcal{B}(K^+_1(1270)\rightarrow \rho^0 K^+)},
\end{align}
and
\begin{align}\label{eq:p2}
&{\mathcal{B}(D^0\to K^+K^-_1(1270), K^-_1(1270)\rightarrow \overline K^{*0}\pi^-)
\over
\mathcal{B}(D^0\to K^+K^-_1(1270), K^-_1(1270)\rightarrow \rho^0 K^-)}\nonumber\\
=&
{\mathcal{B}(D^0\to K^+K^-_1(1270))~\mathcal{B}(K^-_1(1270)\rightarrow \overline K^{*0}\pi^-)
\over
\mathcal{B}(D^0\to K^+K^-_1(1270))~\mathcal{B}(K^-_1(1270)\rightarrow \rho^0 K^-)}
\nonumber\\=&
{\mathcal{B}(K^-_1(1270)\rightarrow \overline K^{*0}\pi^-)
\over
\mathcal{B}(K^-_1(1270)\rightarrow \rho^0 K^-)}.
\end{align}
The equality relation in Eq. (\ref{eq:puzzle}) can then be obtained from Eqs. \eqref{eq:p1} and \eqref{eq:p2}, due to the $CP$ conservation of the strong interaction.

The branching  fractions of the cascade decays involving $K_1(1270)$ are obtained from the fractions by CLEO \cite{Artuso:2012df} shown in Table\ 1 and the data of $\mathcal{B}(D^0\rightarrow K^+K^-\pi^+\pi^-)=(2.42\pm0.12)\times10^{-3}$ \cite{Patrignani:2016xqp},
\begin{align}
&\mathcal{B}_1=\mathcal{B}(D^0\rightarrow K^-K^+_1(1270), K_1^+(1270)\rightarrow K^{*0}\pi^+, K^{*0}\to K^+\pi^-)        =(1.8\pm0.5)\times 10^{-4}   \label{data1},     \\
&\mathcal{B}_2=\mathcal{B}(D^0\rightarrow K^-K^+_1(1270),K_1^+(1270)\rightarrow \rho^0K^+, \rho^0\to \pi^+\pi^-)  =(1.14\pm0.26)\times 10^{-4}        \label{data2},    \\
&\mathcal{B}_3=\mathcal{B}(D^0\rightarrow K^+ K^-_1(1270),K_1^-(1270)\rightarrow \overline{K}^{*0}\pi^-, \overline K^{*0}\to K^-\pi^+) =(2.2\pm1.2)\times 10^{-5}     \label{data3},   \\
&\mathcal{B}_4=\mathcal{B}(D^0\rightarrow K^+K^-_1(1270),K_1^-(1270)\rightarrow \rho^0K^-, \rho^0\to\pi^+\pi^-)           =(1.45\pm0.25)\times 10^{-4}.        \label{data4}
\end{align}
The narrow width approximation indicates
\begin{align}\label{eq:equal}
{\mathcal{B}_1\over \mathcal{B}_2}={\mathcal{B}_3\over \mathcal{B}_4},
\end{align}
while the data in (\ref{data1})$-$(\ref{data4}) give
\begin{align}\label{eq:ratio1}
{\mathcal{B}_1\over \mathcal{B}_2}=1.55\pm0.56,~~~~~\text{and}~~~~~{\mathcal{B}_3\over \mathcal{B}_4}=0.15\pm0.09,
\end{align}
which have large discrepancy with more than 2 standard deviations. The central values of $\mathcal{B}_1/\mathcal{B}_2$ and $\mathcal{B}_3/\mathcal{B}_4$ are even different by a factor of 10. This is the $K_1$ puzzle that the data measured by CLEO are inconsistent with the equality relation of the narrow with approximation.

From Eqs. \eqref{eq:p1} and \eqref{eq:p2}, it can be found that only the strong decays of \ka~are left.
There are some other measurements on the $K_1(1270)$ decays.
It would be useful to compare among the measurements, to give some implications on the solution of the $K_1$ puzzle.
Before the comparison, it is more convenient to define a parameter, $\eta$, describing the ratio of branching fractions of $K_1(1270)\to K^*\pi$ and $K_1(1270)\to \rho K$,
\begin{align}\label{definition}
\eta\equiv\frac{\,\mathcal{B}(K_1(1270)\rightarrow K^*\pi)}{\,\mathcal{B}(K_1(1270)\rightarrow K\rho)},
\end{align}
where the branching fractions are the sums of all the possible charged and neutral final states. For example, $\mathcal{B}(K_1^+(1270)\rightarrow K^*\pi)={3\over2}\mathcal{B}(K_1^+(1270)\rightarrow K^{*0}\pi^+)$ due to the isospin relation of $\mathcal{A}(K_1^+(1270)\rightarrow K^{*0}\pi^+)=-\sqrt{2}\mathcal{A}(K_1^+(1270)\rightarrow K^{*+}\pi^0)$. Similarly, $\mathcal{B}(K_1^+(1270)\rightarrow \rho K)=3\mathcal{B}(K_1^+(1270)\rightarrow \rho^0 K^+)$, $\Gamma_{K^{*0}}={3\over2}\Gamma(K^{*0}\to K^+\pi^-)$. Therefore, the values of $\eta$ obtained from Eq. (\ref{eq:ratio1}) are then
\begin{align}\label{eq:eta12}
&\eta_1={3\over4}{\mathcal{B}_1\over\mathcal{B}_2}=1.16\pm0.42, ~~~~~\text{and}~~~~~\eta_2={3\over4}{\mathcal{B}_3\over\mathcal{B}_4}=0.11\pm0.06.
\end{align}
The $K_1$ puzzle can be taken as the discrepancy between $\eta_1$ and $\eta_2$.

In the following, we discuss on the other measurements which can provide the information on the value of $\eta$.
Except for the singly Cabibbo-suppressed mode of $D^0\to K^+K^-\pi^+\pi^-$, $K_1(1270)\to K^*\pi$ and $\rho K$ are also measured in the Cabibbo-favored $D^0\rightarrow K^-\pi^+\pi^+\pi^-$ decay by BESIII \cite{Ablikim:2017eqz} and LHCb \cite{Aaij:2017kbo}.
With $1.6\times10^4$ signal events of $D^0\rightarrow K^-\pi^+\pi^+\pi^-$ and fixing the mass and width of $K_1(1270)$ as the PDG values, BESIII obtains the branching fractions of \cite{Ablikim:2017eqz}
\begin{align}
&\mathcal{B}_5=\mathcal{B}(D^0\rightarrow \pi^+K^-_1(1270), K_1^-(1270)\rightarrow \overline{K}^{*0}\pi^-, \overline K^{*0}\to K^-\pi^+ ) =(0.07\pm0.02)\%,       \\
&\mathcal{B}_6=\mathcal{B}(D^0\rightarrow \pi^+K^-_1(1270), K_1^-(1270)\rightarrow \rho^0K^-, \rho^0\to \pi^+\pi^- )    =(0.27\pm0.05)\%.
\end{align}
Similarly to Eq. (\ref{eq:eta12}), we have
\begin{align}
\eta_3={3\over4}{\mathcal{B}_5\over\mathcal{B}_6}=0.19\pm0.10,
\end{align}
which is consistent with $\eta_2$.

At the LHCb with even more data of $D^0\rightarrow K^-\pi^+\pi^+\pi^-$ with $9\times10^5$ signal events \cite{Aaij:2017kbo}, more discoveries and higher precisions are obtained. $K_1(1270)\to \rho(1450)K$ is observed and has a relatively large branching fraction. They also find the $D$-wave $K^*\pi$ with a high significance. The interference between amplitudes are considered in \cite{Aaij:2017kbo}. The results of partial fractions are $(96.3\pm1.64\pm6.61)\%$ for $K_1^-(1270)\to \rho^0K^-$, $(27.08\pm0.64\pm2.82)\%$ for $S$-wave $\overline K^{*0}\pi^-$ and $(3.47\pm0.17\pm0.31)\%$ for $D$-wave $\overline K^{*0}\pi^-$.
The phases of the amplitudes of the S-wave and D-wave are $(-172.6\pm1.1\pm6.0)^\circ$ and $(-19.3\pm1.6\pm6.7)^\circ$, respectively.
Then, it is obtained that
\begin{align}
\eta_3'=0.10\pm0.03.
\end{align}
%

The decays of \ka~are also studied in $B^+\rightarrow J/\Psi K^+\pi^+\pi^- $ by Belle \cite{Guler:2010if}. Two amplitude analysis have been performed with the mass and width of \ka~fixed or floated, named as Fit 1 and Fit 2, respectively. The analysis are based on the assumption of \ka~decaying only to $K^*\pi$, $K\rho$, $K\omega$ and $K_0^*(1430)\pi$, and neglect the interference between decay channels.  The results are thus not reliable. We just list them here:

\begin{table}\caption{
Values of observable $\eta$ extracted from different experiments.}
\begin{tabular}{ccccc}

\hline\hline

$~~~~\eta~~~~$             &       Processes        &       Experiments     \\
\hline
$\eta_1=1.16\pm0.42$      &       $D^0\rightarrow K^+K^-\pi^+\pi^- $          &       CLEO \cite{Artuso:2012df}            \\
\hline
$\eta_2=0.11\pm0.06$      &       $D^0\rightarrow K^+K^-\pi^+\pi^- $          &      CLEO \cite{Artuso:2012df}      \\
\hline
$\eta_3=0.19\pm0.10$      &       $D^0\rightarrow K^-\pi^+\pi^+\pi^- $        &      BESIII \cite{Ablikim:2017eqz}\\
\hline
$\eta_3'=0.10\pm0.03$     &       $D^0\rightarrow K^-\pi^+\pi^+\pi^- $        &      LHCb \cite{Aaij:2017kbo} \\
\hline
$\eta_4=0.45\pm0.05$      &       $B^+\rightarrow J/\Psi K^+\pi^+\pi^- $      &     Belle  \cite{Guler:2010if} (Fit 1) \\
\hline
$\eta_4'=0.30\pm0.04$     &       $B^+\rightarrow J/\Psi K^+\pi^+\pi^- $      &       Belle \cite{Guler:2010if} (Fit 2) \\
\hline
$\eta_5=0.38\pm0.13$      &       $K^-p\rightarrow K^-\pi^-\pi^+p $           &     ACCMOR \cite{Daum:1981hb}\\

\hline
\label{eta}
\end{tabular}\\
\end{table}

The values of branching fractions of \ka~decays in PDG are obtained from the $K^-p\to K^-\pi^-\pi^+p$ scattering experiment by the ACCMOR collaboration in 1981 \cite{Daum:1981hb}, with
\begin{align}
&\mathcal{B}(K_1(1270)\to K\rho)=(42\pm6)\%, 
~~~\mathcal{B}(K_1(1270)\to K^*\pi)=(16\pm5)\%,
\end{align}
and thus
\begin{align}
\eta_5=0.38\pm0.13.
\end{align}

 All the values of $\eta$ obtained from different experiments are listed in Table\ 3 for comparison.
 We can find that except $\eta_1$, all the other $\eta$'s indicate a smaller value of $\eta\ll1$, especially $\eta_{2,3,4}=\mathcal{O}(0.1-0.2)$ in $D$ decays.
Thus it is of a large probability that $\eta_1=1.18\pm0.43$ is overestimated.
Due to its large uncertainty, $\eta_1$ can be decreased by about 2 standard deviations to be consistent with other values of $\eta$.

Using the equality relation of Eq. (\ref{eq:equal}) and the measured values of $\mathcal{B}_{1,2,3,4}$ in Eqs. (\ref{data1})-(\ref{data4}), it can be estimated that
\begin{align}\label{eq:B1p}
\mathcal{B}_1'=&\mathcal{B}'(D^0\rightarrow K^-K^+_1(1270), K_1^+(1270)\rightarrow K^{*0}\pi^+, K^{*0}\to K^+\pi^-) \nonumber\\
= &{\mathcal{B}_2\mathcal{B}_3\over\mathcal{B}_4}=(1.7\pm1.1)\times10^{-5},
\end{align}
if $\mathcal{B}_1=(1.8\pm0.5)\times10^{-4}$ was overestimated, or
$\mathcal{B}_2'=\mathcal{B}'(D^0\rightarrow K^-K^+_1(1270),K_1^+(1270)\rightarrow \rho^0K^+,\rho^0\to \pi^+\pi^-) ={\mathcal{B}_1\mathcal{B}_4/\mathcal{B}_3}=(1.2\pm0.8)\times10^{-3}$,
%
if $\mathcal{B}_2=(1.14\pm0.26)\times10^{-4}$ was underestimated. That means, under the equality relation, either $\mathcal{B}_1$ should be reduced to be one-order smaller, or $\mathcal{B}_2$ to be one-order larger. However, with an uncertainty of $20\%$, the measured value of $\mathcal{B}_2$ deviates too much from the central value of $\mathcal{B}_2'$.  Considering the large uncertainty of $\mathcal{B}_{2}'$, the lower bound of $\mathcal{B}_2'$ is close to $\mathcal{B}_2$. Therefore, the true value of $\mathcal{B}(D^0\rightarrow K^-K^+_1(1270),K_1^+(1270)\rightarrow \rho^0K^+,\rho^0\to \pi^+\pi^-)$ would be around $\mathcal{B}_2$. On the contrary, the value of $\mathcal{B}_1'$ deviates from the measured $\mathcal{B}_1$ by about $3\sigma$. It is of large possibility that $\mathcal{B}_1$ is overestimated.

Recall that in the CLEO analysis \cite{Artuso:2012df}, only $K_1^\pm(1270)$ are considered as the $1^+$ states but with $K_1^\pm(1400)$ neglected.
It deserves to test whether \kb~contributes to the overestimation of $\mathcal{B}(D^0\rightarrow K^-K^+_1(1270), K_1^+(1270)\rightarrow K^{*0}\pi^+, K^{*0}\to K^+\pi^-)$.

Note in the end of this section that, we have tested the finite width effect of $K_1(1270)$ in the factorization approach, and find that this effect shifts the branching fractions from the narrow width approximation by less than $10\%$. From Table 3, any uncertainty of the $\eta$'s is larger than $10\%$. Therefore, the finite width effect can be neglected. The narrow width approximation is valid in the discussions.


\section{$D\to K_1(1400)$ transitions}\label{1400}

The contributions from $K_1^\pm(1400)$ in the $D^0\to K^+K^-\pi^+\pi^-$ decay are studied in this section.
The branching fractions of $D^0\rightarrow K^\pm K^\mp_1(1400)(\to \rho K,K^{*}\pi)$ decays are calculated in the factorization approach.
Note that the above processes are kinematically forbidden due to $m_{D^0}< (m_{K_1(1400)}+m_{K})$.
However, the chain decays of $D^0\rightarrow K^\pm K^\mp_1(1400)(\to \rho K,K^{*}\pi)$ can still happen considering the finite width of $K_1(1400)$. From Table\ 2, $m_{K_1(1400)}+m_{K}-m_{D^0}=32\pm7$ MeV $<\Gamma_{K_1(1400)}=174\pm13$ MeV.

The decay constant of axial-vector meson ($A$) and the form factors $D\rightarrow A$ transition are defined as
\begin{align}\label{A}
\langle A(p,\varepsilon)|A_\mu|0\rangle      =& f_Am_A\epsilon^*_\mu,        \nonumber\\
\langle A(p,\varepsilon)|A_\mu|D(p_D)\rangle =& \frac{2}{m_D-m_A}\epsilon_{\mu\nu\alpha\beta}\epsilon^{*\nu}p_D^\alpha p^\beta A^{D\to A}(q^2),   \nonumber\\
\langle A(p,\varepsilon)|V_\mu|D(p_D)\rangle =& -i\bigg\{(m_D-m_A)\epsilon^*_\mu V^{D\to A}_1(q^2) 
                                               -(\epsilon^*\cdot p_D)(p_D+p)_\mu \frac{V^{D\to A}_2(q^2)}{m_D-m_A}   \nonumber\\
                                              &-2m_A\frac{\epsilon^*\cdot p_D}{q^2}q_\mu  \big[V^{D\to A}_3(q^2)
                                               -V^{D\to A}_0(q^2)\big]\bigg\},
\end{align}
in which $q_\mu=(p_D-p)_\mu$.
The decay constant of pseudoscalar meson ($P$) and the form factors of $D\to P$ transition are
\begin{align}\label{B}
\langle P(p) |A_\mu| 0 \rangle &= if_Pp_\mu,        \nonumber\\
\langle P(p) |V_\mu |D(p_D)\rangle &= \bigg( (p_D + p)_\mu - \frac{m_D^2-m^2_P}{q'_\mu} \bigg) F_1^{D\to P}(q'^2) + \frac{m_D^2-m_P^2}{q'^2}q'_\mu F_0^{D\to P}(q'^2),
\end{align}
with $q'_\mu=(p_D - p)_\mu$.
In the factorization approach, the amplitudes of $D^0\rightarrow K^-K^+_1(1400)$ and $D^0\rightarrow K^+K^-_1(1400)$ are expressed as
\begin{align}
\mathcal{M}(D^0\rightarrow K^-K^+_1(1400)) = & -\frac{G_F}{\sqrt{2}}V_{cs}^* V_{us} 
\times\big[2a_1(\mu)\sqrt{q^2}f_{K_1(1400)}F_1^{D\to K}(q^2)\big] (\epsilon^*\cdot p_D),  \label{M1}
\\
\mathcal{M}(D^0\rightarrow K^+K^-_1(1400)) = & \frac{G_F}{\sqrt{2}}V_{cs}^* V_{us} 
\times\big[2a_1(\mu)\sqrt{q^2}f_K(\cos\theta_{K_1} V_0^{D\to K_{1A}}(m^2_K)        \nonumber\\
&  - \sin\theta_{K_1} V_0^{D\to K_{1B}} (m^2_K))\big](\epsilon^*\cdot p_D),\label{M2}
\end{align}
where $\epsilon^*$ is the polarization vector of $K_1(1400)$ and the effective Wilson coefficient
$a_1(\mu)=C_2(\mu) + C_1(\mu)/3$. In this work, we take $\mu=\mu_c=m_c$, so that $a_1(\mu_c)=1.08$ \cite{Li:2012cfa}.
Note that, to consider the finite-width effect \cite{Kamal:1991kg,Cheng:2003bn}, a running mass $\sqrt{q^2}$ for the unstable particle $K_1(1400)$ is considered in Eqs. \eqref{M1} and \eqref{M2}.
According to \cite{Cheng:2003sm}, the form factors of charm decays are parameterized as
\begin{align}\label{eq:ff}
F(q^2)=\frac{F(0)}{1-a(q^2/m^2_D) + b(q^2/m^2_D)^2}.
\end{align}
In this work, the values of form factors of $D\to K_{1A,1B}$ and $K$ are taken from \cite{Cheng:2003sm} in the covariant light-front quark model, as shown in Table \ 4.
The decay constant of $K_1(1400)$ is taken as $139.2^{+41.3}_{-45.6}$ MeV obtained from the $\tau\to K_1(1400)\nu$ decay\footnote{Note that from the $\tau\to K_1(1400)\nu$ decay the decay constant of $K_1(1400)$ is actually obtained as $|f_{K_1(1400)}|=139.2^{+41.3}_{-45.6}$ MeV. Its sign cannot be determined from an individual process. However, in this work our results are independent on the sign of $f_{K_1(1400)}$, since in the factorization approach the decay width of $D^0\to K^-K_1^+(1400)$ is the squared magnitude of the amplitude in Eq. \eqref{M1}. } \cite{Cheng:2010vk}.  The decay constant of $K$ meson is from \cite{Patrignani:2016xqp}.
\begin{table}
\caption{
The form factors of $D\rightarrow K,K_{1A},K_{1B}$ transitions under the parametrization of Eq.\eqref{eq:ff}, taken from the covariant light-front quark model \cite{Cheng:2003sm}.
}
\begin{tabular}{ccccc}

\hline\hline

F                    &      F(0)      &       a          &       b          \\
\hline
$V_0^{DK_{1A}}$      &      0.34      &      1.44        &       0.15       \\
\hline
$V_0^{DK_{1B}}$      &      0.44      &      0.80        &       0.27       \\
\hline
$F_1^{DK}$           &      0.78      &      1.05        &       0.23       \\

\hline
\label{tab:para}
\end{tabular}\\
\end{table}
%

Considering the finite-width effect, the decay widths of the chain decay of $D^0\rightarrow K^\pm K^\mp_1(1400)(\to \rho^0K^\mp \,\,\text{or}\,\,K^{*0}\pi^+,\overline{K}^{*0}\pi^-)$ can be expressed as
\begin{align}\label{eq:BW1}
\Gamma(&D^0\rightarrow K^-K^+_1(1400)(\rightarrow K^{*0}\pi^+))= \int^{(m_D-m_K)^2}_{(m_{K^*}+m_\pi)^2}\frac{dq^2}{\pi}  \nonumber\\
&\Gamma(q^2)(D^0\rightarrow K^-K^+_1(1400))\times\mathcal{B}(K^+_1(1400)\rightarrow K^{*0}\pi^+) \nonumber\\
&\times \frac{\sqrt{q^2}\Gamma(q^2)}{(q^2-M^2)^2-M^2\Gamma^2(q^2)},
\end{align}
\begin{align}
\Gamma(&D^0\rightarrow K^-K^+_1(1400)( \rightarrow \rho^0K^+)) =\int^{(m_D-m_K)^2}_{(m_{\rho}+m_K)^2}\frac{dq^2}{\pi} \nonumber\\
&\Gamma(q^2)(D^0\rightarrow K^-K^+_1(1400))\times\mathcal{B}(K^+_1(1400)\rightarrow \rho^0K^+) \nonumber\\
&\times \frac{\sqrt{q^2}\Gamma(q^2)}{(q^2-M^2)^2-M^2\Gamma^2(q^2)},
\end{align}
\begin{align}
\Gamma(&D^0\rightarrow K^+K^-_1(1400)(\rightarrow \overline{K}^{*0}\pi^-))= \int^{(m_D-m_K)^2}_{(m_{K^*}+m_\pi)^2}\frac{dq^2}{\pi} \nonumber\\ &\Gamma(q^2)(D^0\rightarrow K^+K^-_1(1400))\times\mathcal{B}(K^-_1(1400)\rightarrow \overline{K}^{*0}\pi^-)\nonumber\\
&\times \frac{\sqrt{q^2}\Gamma(q^2)}{(q^2-M^2)^2-M^2\Gamma^2(q^2)},
\end{align}
\begin{align}\label{eq:BW4}
\Gamma(&D^0\rightarrow K^+K^-_1(1400)(\rightarrow \rho^0K^-)) =\int^{(m_D-m_K)^2}_{(m_{\rho}+m_K)^2}\frac{dq^2}{\pi}  \nonumber\\
&\Gamma(q^2)(D^0\rightarrow K^+K^-_1(1400))\times\mathcal{B}(K^-_1(1400)\rightarrow \rho^0K^-) \nonumber\\
&\times \frac{\sqrt{q^2}\Gamma(q^2)}{(q^2-M^2)^2-M^2\Gamma^2(q^2)},
\end{align}
where $\sqrt{q^2}$ is the invariant masses of the $K^*\pi$ and $K\rho$ final states, and $M$ and $\Gamma$ are the mass and width of $K_1(1400)$, respectively.
The $q^2$-dependent width of $K_1(1400)$ is \cite{Blatt:1952ije}:
\begin{align}
\Gamma(q^2)=\Gamma_{K_1(1400)}\frac{M_{K_1(1400)}}{\sqrt{q^2}}\bigg(\frac{p(q^2)}{p(M^2_{K_1(1400)})}\bigg)^3F_R^2(q^2),
\end{align}
in which
\begin{align}
F_R(q^2)=\frac{\sqrt{1+R^2p^2(M^2_{K_1(1400)})}}{\sqrt{1+R^2p^2(q^2)}},
\end{align}
and $p(q^2)=\lambda^{1/2}(q^2,m^2_{1},m^2_2)\big/(2\sqrt{q^2})$, $\lambda(q^2,m^2_{1},m^2_2)=(q^2-(m_1-m_2)^2)(q^2-(m_1+m_2)^2)$, $m_{1,2}$ are the masses of $K^*$ and $\pi$ or $\rho$ and  $K$. The $radius$ of the axial meson is taken as $R$=1.5GeV$^{-1}$ \cite{Kopp:2000gv}.
The branching fractions of $K_1(1400)$ decays are  \cite{Patrignani:2016xqp}
\begin{align}\label{eq:BK1}
&\mathcal{B}(K_1(1400)\to K^*\pi)=(94\pm6)\%,
 ~~~\text{and}~~~ \mathcal{B}(K_1(1400)\to K\rho)=(3.0\pm3.0)\%.
\end{align}

\begin{table}
\caption{
Branching fractions of $D^0\rightarrow K^\pm K^\mp_1(1400)(\to \rho^0K^\pm \,\,\text{or}\,\,K^{*0}\pi^+,\overline{K}^{*0}\pi^-)$ decays with mixing angles $\theta_{K_1}=35^\circ$, $45^\circ$, $55^\circ$ and $60^\circ$.
}
\begin{tabular}{ccccc}
\hline\hline
~~~~~~~~~~~~~~Modes                             & ~~~$\mathcal{B}$ ($\theta_{K_1}=35^\circ$) & ~~~$\mathcal{B}$ ($\theta_{K_1}=45^\circ$) & ~~~$\mathcal{B}$ ($\theta_{K_1}=55^\circ$) &~~~$\mathcal{B}$ ($\theta_{K_1}=60^\circ$) \\
\hline
$K^-K^+_1(1400)(\rightarrow \pi^+K^{*0}(\rightarrow K^+\pi^-))$   &  $(1.3\pm0.9)\times10^{-5}$ &  $(1.3\pm0.9)\times10^{-5}$&$(1.3\pm0.9)\times10^{-5}$ &  $(1.3\pm0.9)\times10^{-5}$\\
\hline
$K^-K^+_1(1400)(\rightarrow K^+\rho^0(\to\pi^+\pi^-))$     &  $(6.5\pm7.8)\times10^{-8}$ &  $(6.5\pm7.8)\times10^{-8}$ & $(6.5\pm7.8)\times10^{-8}$  &  $(6.5\pm7.8)\times10^{-8}$   \\
\hline
$K^+K^-_1(1400)(\rightarrow \pi^-\overline{K}^{*0}(\to K^-\pi^+))$  & $(1.5\pm0.1)\times10^{-8}$ &  $(3.3\pm0.3)\times10^{-8}$  &      $(2.3\pm0.2)\times10^{-7}$ &  $(5.9\pm0.5)\times10^{-7}$  \\
\hline
$K^+K^-_1(1400)(\rightarrow K^-\rho^0(\to\pi^+\pi^-))$               & $(6.8\pm6.8)\times10^{-11}$&  $(1.4\pm1.4)\times10^{-10}$  &    $(1.0\pm1.0)\times10^{-9}$   &    $(2.6\pm2.6)\times10^{-9}$   \\
\hline\hline
\label{brth}
\end{tabular}
\end{table}

%
To calculate the branching fractions, the mixing angle of $\theta_{K_1}$ has to be fixed.
We test the values of $35^\circ$, $45^\circ$, $55^\circ$ and $60^\circ$ which are usually predicted in literatures as shown in the INTRODUCTION.
The numerical results of $D^0\rightarrow K^\pm K^\mp_1(1400)(\to \rho^0K^\pm \,\,\text{or}\,\,K^{*0}\pi^+,\overline{K}^{*0}\pi^-)$ decays are listed in Table\ 5.
The finite width effect allow the $D^0\rightarrow K^\pm K^\mp_1(1400)$ processes to happen.
In principle, the branching fractions depend on the $K_{1}$ mixing angle.
The predictions on $\mathcal{B}(D^0\rightarrow K^-K^+_1(1400)(\to \rho^0K^+ \,\,\text{and}\,\,K^{*0}\pi^+))$ are, nevertheless, invariant for different values of $\theta_{K_1}$, since the mixing angle is involved in the decay constant of $K_1(1400)$ which is however taken as a constant from the $\tau\to K_1(1400)\nu$ decay, seen in Eq. \eqref{M1}.
The branching fractions of the processes associated with $K_{1}(1400)\to K^{*}\pi$ and $\rho K$ differ by about two orders of magnitude, due to the hierarchy of branching fractions of $K_{1}(1400)$ decays in Eq. \eqref{eq:BK1}, and the difference of integral lower limits in Eqs. (\ref{eq:BW1})$-$(\ref{eq:BW4}).
The branching fractions of the $K^{-}K_{1}^{+}(1400)$ modes are larger than those of the $K^{+}K_{1}^{-}(1400)$ modes by two or three orders of magnitude, since the transition form factor of $D\to K_{1}(1400)$ is destructive and suppressed as $(\cos\theta_{K_{1}}V_{0}^{D\to K_{1A}}-\sin\theta_{K_{1}}V_{0}^{D\to K_{1B}})$ with $\theta_{K_{1}}$ in the range between $35^{\circ}$ and $60^{\circ}$, given in Eq. (\ref{M2}).
The uncertainties in our calculation include errors of the width $\Gamma_{K_1(1400)}$, the decay constant $f_{K_1(1400)}$ and the branching fractions of $K_1(1400)\rightarrow K^*\pi$ and $\rho K$ decays.

From Table\ 5, it is found that the branching fraction of $D^0\rightarrow K^-K^+_1(1400)(\rightarrow K^{*0}\pi^+)$ is of the order of $10^{-5}$, same order as our prediction of $\mathcal{B}'(D^0\rightarrow K^-K^+_1(1270)(\rightarrow K^{*0}\pi^+))$ in Eq.~\eqref{eq:B1p}.
The branching fraction of $D^0\to K^-K^+_1(1270),K^+_1(1270)\to K^{*0}\pi^+,K^{*0}\to K^+\pi^-$ is also estimated in the naive factorization in which the width of $K_1(1270)$ is considered as $m_{D^0}-m_{K^\pm}-m_{K_1(1270)} \sim 100$ MeV.  Its value is $(2.19\pm0.88)\times10^{-5}$, and again, being as same order as the branching fraction of $\mathcal{B}(K^-K^+_1(1400)(\rightarrow \pi^+K^{*0}(\rightarrow K^+\pi^-)))=(1.3\pm0.9)\times10^{-5}$.
In order to estimate how large the interference between $D^0\to K^-K^+_1(1270),K^+_1(1270)\to K^{*0}\pi^+,K^{*0}\to K^+\pi^-$ and $D^0\rightarrow K^-K^+_1(1400)(\rightarrow K^{*0}\pi^+)$ could be, we assume that the two chain decays have same phase space, $m_{K_1(1270)}\sim m_{K_1(1400)}$, for simplification, since the amplitudes of the strong decays and their relative phase are unknown.
 Then the total branching fraction of the two chain decays and the maximal interference between them are expected to be $(\sqrt{(2.19\pm0.88)\times10^{-5}}+\sqrt{(1.3\pm0.9)\times10^{-5}})^2=(6.80\pm2.49)\times10^{-5}$  and $2\times\sqrt{(2.19\pm0.88)\times10^{-5}}\times\sqrt{(1.3\pm0.9)\times10^{-5}}=(3.34\pm1.29)\times10^{-5}$, respectively.
Therefore, $D^0\rightarrow K^-K^+_1(1400)(\rightarrow K^{*0}\pi^+)$ might contribute to the overestimation of $\mathcal{B}(D^0\rightarrow K^-K^+_1(1270)(\rightarrow K^{*0}\pi^+))$. The contribution of $K_1(1400)$ cannot be neglected in the experimental analysis.

The estimation of charm decays in the naive factorization approach is not very reliable.
For example, the non-factorizable $W$-exchange diagram $E$ is missed in the above calculation, but is usually large and non-negligible as seen in $D\to PP$  and $PV$ modes \cite{Li:2012cfa,Li:2013xsa,Fusheng:2011tw}.
If more data of $D\to PA$ decays are obtained by experiments, their branching fractions can be calculated in the  factorization-assisted topological amplitude (FAT) approach \cite{Li:2012cfa,Li:2013xsa} in which some global parameters are extracted from data.
More experimental data of $D\to PA$ decays are beneficial to understand the charmed meson decays into axial-vector mesons.

Although $K_1(1400)$ might contribute to the overestimation of $\mathcal{B}_1$, we still cannot conclude whether the $K_1$ puzzle is solved by the consideration of $K_1(1400)$, due to the rough understanding of $D\to PA$ decays.
It has to be tested by the experimental measurements with higher precision, and cross checks from other processes.

\section {Experimental potentials}\label{relation}

The $K_{1}$ puzzle is found in the $D^{0}\to K^{+}K^{-}\pi^{+}\pi^{-}$ decay measured by the CLEO collaboration  \cite{Artuso:2012df}, based on $3\times10^{3}$ signal events.
With such limited data set, the amplitude analysis heavily depends on the model.
Recently, the CLEO data is re-analyzed with improved lineshape parameterizations  \cite{1703.08505}.
With $\mathcal{B}(D^0\rightarrow K^-K^+_1(1270), K_1^+(1270)\rightarrow K^{*0}\pi^+,K^{*0}\to K^{+}\pi^{-})=(1.3\pm0.9)\times10^{-4}$ and $\mathcal{B}(D^0\rightarrow K^-K^+_1(1270),K_1^+(1270)\rightarrow \rho^0K^+,\rho^{0}\to\pi^{+}\pi^{-})  =(2.2\pm0.6)\times10^{-4}$ in \cite{1703.08505}, we can obtain $\eta_{1}'=0.45\pm0.32$, which is smaller than $\eta_{1}=1.18\pm0.43$, but larger than $\eta_{2}=0.11\pm0.06$.
The central value of the branching fraction of $D^0\to K^-K^+_1(1270),K^+_1(1270)\to K^{*0}\pi^+,K^{*0}\to K^+\pi^-$ is larger by one order of magnitude than our prediction in Eq. (\ref{eq:B1p}) based on the equality relation and the previous CLEO result. Besides, it is found a large contribution from $K_{1}(1400)$ in \cite{1703.08505}, with $\mathcal{B}(D^0\rightarrow K^-K^+_1(1400),K^+_1(1400)\rightarrow K^{*0}\pi^+, K^{*0}\to K^{+}\pi^{-})=(3.0\pm1.7)\times10^{-4}$ with its central value larger by one order than our prediction in Table\ 5 under the naive factorization approach, and also larger than $\mathcal{B}(D^0\rightarrow K^-K^+_1(1270), K_1^+(1270)\rightarrow K^{*0}\pi^+,K^{*0}\to K^{+}\pi^{-})=(1.3\pm0.9)\times10^{-4}$. It is a challenge to be understood, since the $K_{1}(1400)$-involved mode should be suppressed by its phase space from the finite-width effect in this kinematically forbidden decay. All the related results are of large uncertainties. The additional four models in \cite{1703.08505} also provide different results. A more precise analysis is required to understand the $D^0\to K^+K^-\pi^+\pi^-$ decay.

LHCb is collecting the data of $D$ decays.
In \cite{Aaij:2017kbo}, LHCb measured the mode of $D^0\to K^-\pi^+\pi^+\pi^-$ with $9\times10^5$ signal events.
Considering the ratio of branching fractions $\mathcal{B}(D^0\to K^+K^-\pi^+\pi^-)/\mathcal{B}(D^0\to K^-\pi^+\pi^+\pi^-)=(3.00\pm0.13)\%$ \cite{Patrignani:2016xqp}, it can be expected that the yields of $D^0\to K^+K^-\pi^+\pi^-$ could be as large as
 $3\times10^4$ at LHCb, since all the final particles of charged kaons or pions are of similar detecting efficiencies. With the much larger data of the $D^0\to K^+K^-\pi^+\pi^-$ decay
at LHCb compared to $3\times10^{3}$ events at CLEO, the equality relation in Eq. (\ref{eq:puzzle}) and the importance of $K_1(1400)$ could be tested.

The equality relation in \eqref{eq:puzzle} is given by the ratios between the same weak decays, such as $D^0\to K^- K_1^+(1270), K_1^+(1270)\to K^{*0}\pi^+$ v.s. $D^0\to K^- K_1^+(1270), K_1^+(1270)\to \rho^0 K^+$. In this way, the weak decay parts are cancelled in the narrow width approximation. On the other hand, the equality relation can also be expressed as
%
%
%
\begin{align}\label{eq:puzzle2}
&\frac{\Gamma(D^0\to K^-K^+_1(1270)(\rightarrow \pi^+K^{*0}(\to K^+\pi^-)))}{\Gamma(D^0\to K^-K^+_1(1270)(\rightarrow K^+\rho^0(\to \pi^+\pi^-)))}\nonumber\\
=&\frac{\Gamma(D^0\to K^+K^-_1(1270)(\rightarrow \pi^-\overline{K}^{*0}(\to K^-\pi^+)))}{\Gamma(D^0\to K^+K^-_1(1270)(\rightarrow K^-\rho^0(\to\pi^+\pi^-)))}.
\end{align}
Experimental measurements can use the equality relation in the formula as either Eq. \eqref{eq:puzzle} or Eq. \eqref{eq:puzzle2}.

Except for testing the equality relation in the $D^0\to K^+K^-\pi^+\pi^-$ decay, it is also helpful to measure the ratios or test the relations in other four-body $D$ decays, such as $D^0\rightarrow K^0_SK^0_S\pi^+\pi^-$, $D^+\rightarrow K^0_S\pi^+\pi^0\pi^0$, $D^+_s\rightarrow K^0_S\pi^+\pi^+\pi^-$, etc.
The $K_1(1270)$ resonance exists in such processes.
 All of the ratios or relations are listed in Tables\ 6 and\ 7, for the Cabibbo-favored and singly Cabibbo-suppressed modes, respectively.
The ratios are given by the $\eta$ parameter defined in Eq.~\eqref{definition}, with the factors from the isospin analysis of strong decays of $K_1(1270)$, $\rho$ and $K^*$.
Any ratio in Tables\ 6 and\ 7 can be measured to be compared with those in Table\ 3. More measurements on $\eta$ will help to solve the $K_1$ puzzle.

Note that all the processes listed in Tables\ 6 and\ 7 satisfy that $m_{D_{(s)}}-(m_{K_1(1270)}+m_{\pi,K})\gtrsim\Gamma_{K_1(1270)}$, so that the narrow width approximation is still valid in these processes. Besides, in the $K_S^0$ involved modes in Table\ 6, the doubly Cabibbo-suppressed amplitudes are neglected due to their smallness.

In Tables\ 6 and\ 7, we only list the observables associated with $K_1(1270)\to K^*\pi$ and $\rho K$, which are relevant to the $K_1$ puzzle. Actually, the ratios could be between any decay modes of $K_1(1270)$, for example, the fractions between the $D$-wave and $S$-wave widths of $K_1(1270)\to K^*\pi$ and $\rho K$. More precise measurements on \ka~decays are helpful for the determination of the mixing angle $\theta_{K_1}$ \cite{Suzuki:1993yc,Divotgey:2013jba,1705.03144,Tayduganov:2011ui}

Some of  the processes in Tables\ 6 and\ 7 are more preferred in the experimental measurements. Firstly, the branching fractions of the Cabibbo-favored modes are usually large,
and hence easier to be measured.
In the decay of $D_s^+\to K^+K_S^0\pi^+\pi^-$ with a large branching fraction of $(1.03\pm0.10)\%$ \cite{Patrignani:2016xqp}, there are four $K_1(1270)$ related processes. Thus the equality relation can be directly tested with the ratios in $D_s^+\to K_S^0K_1^+(1270)$ and $D_s^+\to K^+\overline K_1^0(1270)$. The $D^0\to K_S^0\pi^+\pi^-\pi^0$ decay, with $\mathcal{B}=(5.1\pm0.6)\%$, also has four $K_1(1270)$ related processes to test the equality relation. The observables in Tables\ 6 and\ 7 can be measured and tested by BESIII, Belle II and LHCb in the near future.

\section{conclusions}\label{con}

Charmed meson decays can provide much useful information about strange axial-vector mesons.
In this work, it is found that the data of $K_1(1270)$ related processes in the $D^0\to K^+K^-\pi^+\pi^-$ mode are inconsistent with the equality relation under the narrow width approximation and $CP$ conservation of strong decays.
The ratio between $\mathcal{B}(D^0\rightarrow K^-K^+_1(1270)(\to \pi^+K^{*0}(\to K^+\pi^-)))$ and $\mathcal{B}(D^0\to K^-K^+_1(1270)(\rightarrow K^+\rho^0(\to \pi^+\pi^-)))$, with a value of $1.58\pm0.57$, deviates by about 2$\sigma$ from the one between $\mathcal{B}(D^0\to K^+K^-_1(1270)(\rightarrow \pi^-\overline{K}^{*0}(\to K^-\pi^+)))$ and $\mathcal{B}(D^0\to K^+K^-_1(1270)(\rightarrow K^-\rho^0(\to\pi^+\pi^-)))$ with a value of $0.15\pm0.09$.
In the amplitude analysis by CLEO of the above measurement, $K_1(1400)$ was neglected.
We calculate the branching fractions of the $D^0\rightarrow K^\pm_1(1400)(\to \rho^0K^\pm \,\,\text{or}\,\,K^{*0}\pi^+,\overline{K}^{*0}\pi^-)K^\mp$ modes using the factorization approach considering the finite-width effect. It is found that the branching fraction of $D^0\rightarrow K^-K^+_1(1400)(\to \pi^+K^{*0}(\to K^+\pi^-))$ is comparable to $D^0\rightarrow K^-K^+_1(1270)(\to \pi^+K^{*0}(\to K^+\pi^-))$, and hence might contribute to the overestimation of the latter process. Thus $K_1(1400)$ could not be neglected in the analysis.
In addition, some relations in other $D$ decay modes to study $K_1(1270)$ decays are proposed to be tested by BESIII, Belle (II) and LHCb.

\begin{center}
\begin{table}\caption{ 
The relations of the branching fractions of the Cabbibo-favored cascade decays listed in the table, in which $\eta$ is defined by Eq.~\eqref{definition}.
}
\newcommand{\tabincell}[2]{\begin{tabular}{@{}#1@{}}#2\end{tabular}}
\begin{tabular}{ccccc}

\hline\hline

Four-body decays        &       ~~~~~~~~~~~~~~~~~~~~~Resonant processes     &      ~~~Relations    \\
\hline
$D^0\rightarrow K^+K^-\pi^+\pi^-$   &
\tabincell{l}{
$\mathcal{B}_{11}=\mathcal{B}(D^0\rightarrow K^+_1(1270)K^-,~K^+_1\rightarrow K^{*0}\pi^+,~K^{*0}\rightarrow K^+\pi^-)$\\
$\mathcal{B}_{12}=\mathcal{B}(D^0\rightarrow K^+_1(1270)K^-,~K^+_1\rightarrow \rho^0K^+,~\rho^0\rightarrow\pi^+\pi^-)$\\
$\mathcal{B}_{13}=\mathcal{B}(D^0\rightarrow K^-_1(1270)K^+,~K^-_1\rightarrow\overline{K}^{*0}\pi^-,~\overline{K}^{*0}\rightarrow K^-\pi^+)$\\
$\mathcal{B}_{14}=\mathcal{B}(D^0\rightarrow K^-_1(1270)K^+,~K^-_1\rightarrow\rho^0K^-,~\rho^0\rightarrow\pi^+\pi^-)$}
& \tabincell{c}{$~\mathcal{B}_{11}/\mathcal{B}_{12}=4\eta/3,$\\$\mathcal{B}_{13}/\mathcal{B}_{14}=4\eta/3$} \\
\hline
$D^0\rightarrow K^0_SK^0_S\pi^+\pi^-$     &
\tabincell{l}{
$\mathcal{B}_{21}=\mathcal{B}(D^0\rightarrow K^0_1(1270)K_S^0,~K^0_1\rightarrow K^{*+}\pi^-,~K^{*+}\rightarrow K^0_S\pi^+)$\\
$\mathcal{B}_{22}=\mathcal{B}(D^0\rightarrow K^0_1(1270)K_S^0,~K^0_1\rightarrow \rho^0K_S^0,~\rho^0\rightarrow\pi^+\pi^-)$\\
$\mathcal{B}_{23}=\mathcal{B}(D^0\rightarrow \overline{K}^0_1(1270)K_S^0,~\overline{K}^0_1\rightarrow K^{*-}\pi^+,~K^{*-}\rightarrow K^0_S\pi^-)$\\
$\mathcal{B}_{24}=\mathcal{B}(D^0\rightarrow \overline{K}^0_1(1270)K_S^0,~\overline{K}^0_1\rightarrow\rho^0 K_S^0,~\rho^0\rightarrow\pi^+\pi^-)$
}   &
\tabincell{c}{$~\mathcal{B}_{21}/\mathcal{B}_{22}=4\eta/3,$\\$\mathcal{B}_{23}/\mathcal{B}_{24}=4\eta/3$} \\
\hline
$D^0\rightarrow K^-K^0_S\pi^+\pi^0$     &
\tabincell{l}{
$\mathcal{B}_{31}=\mathcal{B}(D^0\rightarrow K^+_1(1270)K^-,~K^+_1\rightarrow K^{*0}\pi^+,~K^{*0}\rightarrow K^0_S\pi^0)$\\
$\mathcal{B}_{32}=\mathcal{B}(D^0\rightarrow K^+_1(1270)K^-,~K^+_1\rightarrow \rho^+ K_S^0,~\rho^+\rightarrow\pi^+\pi^0)$\\
$\mathcal{B}_{33}=\mathcal{B}(D^0\rightarrow \overline{K}^0_1(1270)K_S^0,~\overline{K}^0_1\rightarrow K^{*-}\pi^+,~K^{*-}\rightarrow K^-\pi^0)$\\
$\mathcal{B}_{34}=\mathcal{B}(D^0\rightarrow \overline{K}^0_1(1270)K_S^0,~\overline{K}^0_1\rightarrow \rho^+K^-,~\rho^+\rightarrow\pi^+\pi^0)$
}   &
\tabincell{c}{$~\mathcal{B}_{31}/\mathcal{B}_{32}=\eta/3,$\\$\mathcal{B}_{33}/\mathcal{B}_{34}=\eta/3$} \\
\hline
$D^0\rightarrow K^+K^0_S\pi^-\pi^0$     &
\tabincell{l}{
$\mathcal{B}_{41}=\mathcal{B}(D^0\rightarrow K^-_1(1270)K^+,~K^-_1\rightarrow\overline{K}^{*0}\pi^-,~\overline{K}^{*0}\rightarrow K^0_S\pi^0)$\\
$\mathcal{B}_{42}=\mathcal{B}(D^0\rightarrow K^-_1(1270)K^+,~K^-_1\rightarrow\rho^- K_S^0,~\rho^-\rightarrow \pi^-\pi^0)$\\
$\mathcal{B}_{43}=\mathcal{B}(D^0\rightarrow K^0_1(1270)K_S^0,~K^0_1\rightarrow K^{*+}\pi^-,~K^{*+}\rightarrow K^+\pi^0)$\\
$\mathcal{B}_{44}=\mathcal{B}(D^0\rightarrow K^0_1(1270)K_S^0,~K^0_1\rightarrow\rho^- K^+,~\rho^-\rightarrow \pi^-\pi^0)$
}   &
\tabincell{c}{$~\mathcal{B}_{41}/\mathcal{B}_{42}=\eta/3,$\\$\mathcal{B}_{43}/\mathcal{B}_{44}=\eta/3$} \\
\hline
$D^+\rightarrow K^+K^0_S\pi^+\pi^-$     &
\tabincell{l}{
$\mathcal{B}_{51}=\mathcal{B}(D^+\rightarrow K^+_1(1270)K_S^0,~K^+_1\rightarrow K^{*0}\pi^+,~K^{*0}\rightarrow K^+\pi^-)$\\
$\mathcal{B}_{52}=\mathcal{B}(D^+\rightarrow K^+_1(1270)K_S^0,~K^+_1\rightarrow \rho^0K^+,~\rho^0\rightarrow\pi^+\pi^-)$\\
$\mathcal{B}_{53}=\mathcal{B}(D^+\rightarrow \overline{K}^0_1(1270)K^+,~\overline{K}^0_1\rightarrow K^{*-}\pi^+,~K^{*-}\rightarrow K^0_S\pi^-)$\\
$\mathcal{B}_{54}=\mathcal{B}(D^+\rightarrow \overline{K}^0_1(1270)K^+,~\overline{K}^0_1\rightarrow\rho^0 K_S^0,~\rho^0\rightarrow\pi^+\pi^-)$
}   &
\tabincell{c}{$~\mathcal{B}_{51}/\mathcal{B}_{52}=4\eta/3,$\\$\mathcal{B}_{53}/\mathcal{B}_{54}=4\eta/3$} \\
\hline
$D^+\rightarrow K^0_SK^0_S\pi^+\pi^0$       &
\tabincell{l}{
$\mathcal{B}_{61}=\mathcal{B}(D^+\rightarrow K^+_1(1270)K_S^0,~K^+_1\rightarrow K^{*0}\pi^+,~K^{*0}\rightarrow K_S^0\pi^0)$\\
$\mathcal{B}_{62}=\mathcal{B}(D^+\rightarrow K^+_1(1270)K_S^0,~K^+_1\rightarrow \rho^+ K_S^0,~\rho^+\rightarrow\pi^+\pi^0)$\\
}   &
$\mathcal{B}_{61}/\mathcal{B}_{62}=\eta/3$\\
\hline
$D^+\rightarrow K^+K^-\pi^+\pi^0$       &
\tabincell{l}{
$\mathcal{B}_{71}=\mathcal{B}(D^+\rightarrow \overline{K}^0_1(1270)K^+,~\overline{K}^0_1\rightarrow K^{*-}\pi^+,~K^{*-}\rightarrow K^-\pi^0)$\\
$\mathcal{B}_{72}=\mathcal{B}(D^+\rightarrow \overline{K}^0_1(1270)K^+,~\overline{K}^0_1\rightarrow\rho^+K^-,~\rho^+\rightarrow\pi^+\pi^0)$\\
}   &
$\mathcal{B}_{71}/\mathcal{B}_{72}=\eta/3$\\
\hline

$D^+_s\rightarrow K^+\pi^+\pi^-\pi^0$     &
\tabincell{l}{
$\mathcal{B}_{81}=\mathcal{B}(D_s^+\rightarrow K^0_1(1270)\pi^+,~K^0_1\rightarrow K^{*+}\pi^-,~K^{*+}\rightarrow K^+\pi^0)$\\
$\mathcal{B}_{82}=\mathcal{B}(D_s^+\rightarrow K^0_1(1270)\pi^+,~K^0_1\rightarrow\rho^- K^+,~\rho^-\rightarrow \pi^-\pi^0)$\\
$\mathcal{B}_{83}=\mathcal{B}(D_s^+\rightarrow K^+_1(1270)\pi^0,~K^+_1\rightarrow K^{*0}\pi^+,~K^{*0}\rightarrow K^+\pi^-)$\\
$\mathcal{B}_{84}=\mathcal{B}(D_s^+\rightarrow K^+_1(1270)\pi^0,~K^+_1\rightarrow\rho^0K^+,~\rho^0\rightarrow\pi^+\pi^-)$
}   &
\tabincell{c}{$\mathcal{B}_{81}/\mathcal{B}_{82}=\eta/3,$\\$\mathcal{B}_{83}/\mathcal{B}_{84}=4\eta/3$}\\
\hline
$D^+_s\rightarrow K^0_S\pi^+\pi^+\pi^-$     &
\tabincell{l}{
$\mathcal{B}_{91}=\mathcal{B}(D_s^+\rightarrow K^0_1(1270)\pi^+,~K^0_1\rightarrow K^{*+}\pi^-,~K^{*+}\rightarrow K_S^0\pi^+)$\\
$\mathcal{B}_{92}=\mathcal{B}(D_s^+\rightarrow K^0_1(1270)\pi^+,~K^0_1\rightarrow\rho^0K_S^0,~\rho^0\rightarrow\pi^+\pi^-)$\\
}   &
$\mathcal{B}_{91}/\mathcal{B}_{92}=4\eta/3$\\
\hline
$D^+_s\rightarrow K^0_S\pi^+\pi^0\pi^0$     &
\tabincell{l}{
$\mathcal{B}_{101}=\mathcal{B}(D_s^+\rightarrow K^+_1(1270)\pi^0,~K^+_1\rightarrow K^{*0}\pi^+,~K^{*0}\rightarrow K_S^0\pi^0)$\\
$\mathcal{B}_{102}=\mathcal{B}(D_s^+\rightarrow K^+_1(1270)\pi^0,~K^+_1\rightarrow\rho^+ K_S^0,~\rho^+\rightarrow \pi^+\pi^0)$\\
}   &
$\mathcal{B}_{101}/\mathcal{B}_{102}=\eta/3$\\

\hline\hline
\label{tab:CF}
\end{tabular}
\end{table}
\end{center}

\newpage

\begin{center}
\begin{table}\caption{ 
Same as Table\ 6 but for singly Cabibbo-suppressed modes.
}
\newcommand{\tabincell}[2]{\begin{tabular}{@{}#1@{}}#2\end{tabular}}
\begin{tabular}{ccccc}

\hline\hline

Four-body decays        &      ~~~~~~~~~~~~~~~~~~~~~Resonant processes     &      ~~~Relations    \\
\hline
$D^0\rightarrow K^0_S\pi^+\pi^-\pi^0$
&
\tabincell{l}{$\mathcal{B}_{11}=\mathcal{B}(D^0\rightarrow K^-_1(1270)\pi^+,~K^-_1\rightarrow \overline{K}^{*0}\pi^-,~\overline{K}^{*0}\rightarrow K^0_S\pi^0)$  \\
$\mathcal{B}_{12}=\mathcal{B}(D^0\rightarrow K^-_1(1270)\pi^+,~K^-_1\rightarrow\rho^- K_S^0,~\rho^-\rightarrow \pi^-\pi^0)$  \\
$\mathcal{B}_{13}=\mathcal{B}(D^0\rightarrow \overline{K}^0_1(1270)\pi^0,~\overline{K}^0_1\rightarrow K^{*-}\pi^+,~K^{*-}\rightarrow K^0_S\pi^- )$  \\
$\mathcal{B}_{14}=\mathcal{B}(D^0\rightarrow \overline{K}^0_1(1270)\pi^0,~\overline{K}^0_1\rightarrow\rho^0 K^0_S,~\rho^0\rightarrow \pi^+\pi^-)$   \\      }
&
\tabincell{l}{$\mathcal{B}_{11}/\mathcal{B}_{12}=\eta/3,~$    \\
$\mathcal{B}_{13}/\mathcal{B}_{14}=4\eta/3$}
\\
\hline
$D^0\rightarrow K^-\pi^+\pi^+\pi^-$     &
\tabincell{l}{
$\mathcal{B}_{21}=\mathcal{B}(D^0\rightarrow K^-_1(1270)\pi^+,~K^-_1\rightarrow\overline{K}^{*0}\pi^-,~\overline{K}^{*0}\rightarrow K^-\pi^+)$\\
$\mathcal{B}_{22}=\mathcal{B}(D^0\rightarrow K^-_1(1270)\pi^+,~K^-_1\rightarrow\rho^0K^-,~\rho^0\rightarrow \pi^+\pi^-)$\\
}   &
$\mathcal{B}_{21}/\mathcal{B}_{22}=4\eta/3$\\
\hline
$D^0\rightarrow K^-\pi^+\pi^0\pi^0$     &
\tabincell{l}{
$\mathcal{B}_{31}=\mathcal{B}(D^0\rightarrow \overline{K}^0_1(1270)\pi^0,~\overline{K}^0_1\rightarrow K^{*-}\pi^+,~K^{*-}\rightarrow K^-\pi^0)$\\
$\mathcal{B}_{32}=\mathcal{B}(D^0\rightarrow \overline{K}^0_1(1270)\pi^0,~\overline{K}^0_1\rightarrow \rho^+K^-,~\rho^+\rightarrow\pi^+\pi^0)$\\
}   &
$\mathcal{B}_{31}/\mathcal{B}_{32}=\eta/3$\\
\hline
$D^+\rightarrow K^0_S\pi^+\pi^+\pi^-$       &
\tabincell{l}{
$\mathcal{B}_{43}=\mathcal{B}(D^+\rightarrow \overline{K}^0_1(1270)\pi^+,~\overline{K}^0_1\rightarrow K^{*-}\pi^+,~K^{*-}\rightarrow K_S^0\pi^-)$\\
$\mathcal{B}_{44}=\mathcal{B}(D^+\rightarrow \overline{K}^0_1(1270)\pi^+,~\overline{K}^0_1\rightarrow\rho^0 K_S^0,~\rho^0\rightarrow\pi^+\pi^-)$
}   &
$\mathcal{B}_{41}/\mathcal{B}_{42}=4\eta/3$\\
%
%
\hline
$D^+\rightarrow K^-\pi^+\pi^+\pi^0$       &
\tabincell{l}{
$\mathcal{B}_{51}=\mathcal{B}(D^+\rightarrow \overline{K}^0_1(1270)\pi^+,~\overline{K}^0_1\rightarrow K^{*-}\pi^+,~K^{*-}\rightarrow K^-\pi^0)$\\
$\mathcal{B}_{52}=\mathcal{B}(D^+\rightarrow \overline{K}^0_1(1270)\pi^+,~\overline{K}^0_1\rightarrow\rho^+K^-,~\rho^+\rightarrow\pi^+\pi^0)$\\
}   &
$\mathcal{B}_{51}/\mathcal{B}_{52}=\eta/3$\\
\hline
$D^+_s\rightarrow K^+K^0_S\pi^+\pi^-$     &
\tabincell{l}{
$\mathcal{B}_{61}=\mathcal{B}(D_s^+\rightarrow K^+_1(1270)K_S^0,~K^+_1\rightarrow K^{*0}\pi^+,~K^{*0}\rightarrow K^+\pi^- )$\\
$\mathcal{B}_{62}=\mathcal{B}(D_s^+\rightarrow K^+_1(1270)K_S^0,~K^+_1\rightarrow \rho^0K^+,~\rho^0\rightarrow\pi^+\pi^-)$\\
$\mathcal{B}_{63}=\mathcal{B}(D_s^+\rightarrow \overline{K}^0_1(1270)K^+,~\overline{K}^0_1\rightarrow K^{*-}\pi^+,~K^{*-}\rightarrow K^0_S\pi^-)$\\
$\mathcal{B}_{64}=\mathcal{B}(D_s^+\rightarrow \overline{K}^0_1(1270)K^+,~\overline{K}^0_1\rightarrow\rho^0 K_S^0,~\rho^0\rightarrow\pi^+\pi^-)$
}   &
\tabincell{c}{$~\mathcal{B}_{61}/\mathcal{B}_{62}=4\eta/3,$\\$\mathcal{B}_{63}/\mathcal{B}_{64}=4\eta/3$} \\
\hline
$D^+_s\rightarrow K^0_SK_S^0\pi^+\pi^0$     &
\tabincell{l}{
$\mathcal{B}_{71}=\mathcal{B}(D_s^+\rightarrow K^+_1(1270)K_S^0,~K^+_1\rightarrow K^{*0}\pi^+,~K^{*0}\rightarrow K^0_S\pi^0)$\\
$\mathcal{B}_{72}=\mathcal{B}(D_s^+\rightarrow K^+_1(1270)K_S^0,~K^+_1\rightarrow\rho^+ K_S^0,~\rho^+\rightarrow\pi^+\pi^0)$\\
}   &
$\mathcal{B}_{71}/\mathcal{B}_{72}=\eta/3$\\
\hline
$D^+_s\rightarrow K^+K^-\pi^+\pi^0$     &
\tabincell{l}{
$\mathcal{B}_{81}=\mathcal{B}(D_s^+\rightarrow \overline{K}^0_1(1270)K^+,\overline{K}^0_1\rightarrow K^{*-}\pi^+,~K^{*-}\rightarrow K^-\pi^0)$\\
$\mathcal{B}_{82}=\mathcal{B}(D_s^+\rightarrow \overline{K}^0_1(1270)K^+,\overline{K}^0_1\rightarrow\rho^+K^-,~\rho^+\rightarrow\pi^+\pi^0)$\\
}   &
$\mathcal{B}_{81}/\mathcal{B}_{82}=\eta/3$\\

\hline\hline
\label{tab:SCS}
\end{tabular}
\end{table}
\end{center}


\renewcommand\refname{{\normalsize\bf References:}}   
\vspace{7mm}
\linespread{1.5}

\end{document}